\documentclass[a4paper,11pt]{article}
\usepackage{jheppub}
\usepackage{lineno}
\usepackage[italicdiff]{physics}
\usepackage{bm,amsfonts,siunitx,slashed,mathtools,amssymb,here}

\preprint{YITP-23-29}
\arxivnumber{2303.05481}

\title{\boldmath Monte Carlo study of Schwinger model \\ without the sign problem}

\author{Hiroki Ohata}
\affiliation{Yukawa Institute for Theoretical Physics, Kyoto University,\\
Sakyo-ku, Kyoto 606-8502, Japan}
\emailAdd{hiroki.ohata@yukawa.kyoto-u.ac.jp}

\abstract{
Monte Carlo study of the Schwinger model (quantum electrodynamics in one spatial dimension) with a topological $\theta$ term is very difficult due to the sign problem in the conventional lattice formulation.
In this paper, we point out that this problem can be circumvented by utilizing the lattice formulation of the bosonized Schwinger model, initially invented by Bender et al. in 1985.
After conducting a detailed review of their lattice formulation, we explicitly validate its correctness through detailed comparisons with analytical and previous numerical results at $\theta = 0$.
We also obtain the $\theta$ dependence of the chiral condensate and successfully reproduce the mass perturbation result for small fermion masses $m / g \lesssim 0.125$.
As an application, we perform a precise calculation of the string tension and quantitatively reveal the confining properties in the Schwigner model at finite temperature and $\theta$ region for the first time.
In particular, we find that the string tension is negative for noninteger probe charges around $\theta = \pi$ at low temperatures.
}

\begin{document}
\maketitle
\flushbottom

\section{Introduction}
Monte Carlo study of lattice quantum chromodynamics (QCD) is established as the most reliable method to investigate the static properties of the strong interaction.
However, in certain situations such as with a topological $\theta$ term or at finite density,
the Euclidean action of QCD can become complex, 
and the stochastic estimation of the Euclidean path-integral is no longer possible.
This is the sign problem in QCD.
Since the same problem appears in various fields of physics, 
the search for a new method to overcome the sign problem is of significant importance. 

The Schwinger model (quantum electrodynamics in one spatial dimension)~\cite{Schwinger:1962tp} has been often utilized as a testing ground 
 of methods aimed at overcoming the sign problem.
There are several reasons.
First, the Schwinger model shares many low-energy phenomena 
with QCD, thereby the model offers valuable insights into the strong interaction.
Secondly, owing to its low dimensionality, 
the Schwinger model can be well investigated analytically using bosonization~\cite{Coleman:1974bu, Mandelstam:1975hb, Coleman:1975pw, Coleman:1976uz, Fischler:1978ms, Manton:1985jm, Iso:1988zi, Hetrick:1988yg, Smilga:1992hx, Smilga:1996pi} and even exactly solvable when the fermion is massless~\cite{Schwinger:1962tp, Manton:1985jm, Iso:1988zi, Hetrick:1988yg, Sachs:1991en}.
This enables us to check our numerical results to some extent.
Also, the Schwinger model can be transformed into a spin system by integrating out the gauge fields and using the Jordan--Wigner transformation.
The dimension of the resulting spin Hamiltonian is finite, albeit exponentially large.
Recently, many approaches have been applied to the Schwinger model, 
including the tensor network method~\cite{Byrnes:2002nv, Banuls:2013jaa, Buyens:2013yza, Shimizu:2014uva, Shimizu:2014fsa, Buyens:2014pga, Banuls:2015sta, Buyens:2015tea, Banuls:2016lkq, Buyens:2016ecr, Banuls:2016gid, Buyens:2017crb, Funcke:2019zna, Ercolessi:2017jbi}, 
the quantum computing~\cite{Kuhn:2014rha, Zache:2018cqq, Kokail:2018eiw, Magnifico:2019kyj, Chakraborty:2020uhf, Honda:2021aum, Thompson:2021eze, Honda:2021ovk, Halimeh:2022pkw, Xie:2022jgj}, 
the dual formulation~\cite{Gattringer:2015nea, Goschl:2017kml}, 
and the Lefschetz thimble method~\cite{Tanizaki:2016xcu, Alexandru:2018ngw}.
Among them, the tensor network method based on the spin Hamiltonian formulation has achieved unparalleled success so far.

In 1985, well before these studies, Bender, Rothe, and Rothe developed the lattice formulation of the bosonized Schwinger model and calculated the static potential at $\theta = 0$ by evaluating its ground state energy in the presence of static probe charges~\cite{Bender:1984qg}.
A notable feature of their lattice formulation was their method of addressing the normal ordering, which arises in the bosonized Hamiltonian for regularization.
Their study provided a distinct lattice formulation of the Schwinger model, differing from the conventional ones, such that using the Kogut--Susskind formulation~\cite{Kogut:1974ag}.
Despite its significance, their paper has not received much attention to date for some reason.

In this paper, we propose to utilize the lattice bosonized Schwinger model as a method to circumvent the sign problem in the Schwinger model. 
In this formulation, the Euclidean action is real even with a $\theta$ term.
Hence the sign problem does not emerge.
After conducting a detailed review of the lattice formulation, we explicitly validate its correctness by reproducing previous analytical and numerical results.
As an application, we perform an extensive calculation of the string tension and quantitatively reveal the confining properties in the Schinger model at finite temperature and $\theta$ for the first time.
The present method is quite simple and the same idea could be straightforwardly applied to a wide variety of fermionic models in one spatial dimension.

This paper is organized as follows.
In section~\ref{sec:formulation}, we provide a comprehensive review of the lattice bosonized Schwinger model. 
In particular, we explicitly document the correspondence with the original Schwinger model's bare fermion mass, which was absent in Ref.~\cite{Bender:1984qg}.
In section~\ref{sec:vertification}, we verify the lattice formulation by reproducing previous analytical and numerical results.
In section~\ref{sec:confinement}, we explain our method to calculate the string tension and perform an extensive calculation of the string tension at finite temperature and $\theta$.
Section~\ref{sec:summary} is devoted to summary and future study.

\section{Lattice bosonized Schwinger model} \label{sec:formulation}
In this section, we conduct a comprehensive review of the lattice bosonized Schwinger model of Bender et al. in 1985~\cite{Bender:1984qg}.
The Euclidean action of the (original) Schwinger model with a $\theta$ term reads
\begin{equation}
S_E = \int d^2 x\, \overline{\psi}\qty(\slashed{\partial} + g \slashed{A} + m)\psi + \frac{1}{4} F_{\mu\nu} F_{\mu\nu} + i\theta \frac{g}{4\pi} \epsilon_{\mu\nu}F_{\mu\nu}.
\end{equation}
Here $F_{\mu\nu} = \partial_\mu A_\nu - \partial_\nu A_\mu$ is the field strength of the $\mathrm{U}(1)$ gauge field $A_\mu$, $\psi$ the Dirac fermion, $g$ the dimensionful gauge coupling, and $m$ the fermion mass.
After bosonization and integrating out the gauge fields, the Hamiltonian of the bosonized Schwinger model reads~\cite{Coleman:1975pw}
\begin{equation}
H = \int d x\, \frac{1}{2}\pi^2 + \frac{1}{2} \qty(\partial_x \phi)^2 + \frac{g^2}{2\pi} \qty(\phi + \frac{\theta}{2 \sqrt{\pi}})^2
- \frac{e^\gamma}{2 \pi^{3/2}} m g \mathcal{N}_{g / \sqrt{\pi}} \cos(2\sqrt{\pi} \phi). \label{eq:continuum_hamiltonian}
\end{equation}
Here $\pi$ is the conjugate momentum, $\gamma$ Euler's constant,
and $\mathcal{N}_{g / \sqrt{\pi}}$ denotes the normal ordering with respect to the boson mass $g / \sqrt{\pi}$. 
\footnote{The prefactor of the cosine term depends on the choice of scale used to define the normal ordering. 
If the normal ordering is taken with respect to the fermion mass, rather than the boson mass, the term takes the form $\frac{e^\gamma}{2\pi} m^2 \mathcal{N}_m \cos(2\sqrt{\pi} \phi)$. 
The equivalence between these two is ensured by Eq.~(\ref{eq:renormal}). 
In Ref.~\cite{Bender:1984qg}, the cosine term was introduced as $\frac{M^2}{4\pi} \cos(2\sqrt{\pi} \phi)$ without explicitly specifying the scale of the normal ordering.}
It is evident from this form that the $\theta$ term is irrelevant at $m = 0$.
For the path-integral formulation of this model on a lattice, the normal ordering appearing in the cosine term must be removed properly.

In the seminal paper on bosonization in 1975~\cite{Coleman:1974bu}, using Wick's theorem, Coleman showed that the normal ordering can be removed as
\begin{equation}
\mathcal{N}_{\mu} \exp(i \beta \phi) = \exp{\frac{\beta^2}{2} \Delta \qty(x = 0; \mu)} \exp(i \beta \phi), \label{eq:NO}
\end{equation}
where $\Delta \qty(x; \mu)$ is the Feynman propagator for the scalar field of mass $\mu$, and $\beta$ is an arbitrary real number.
The Feynman propagator is divergent at the origin in the continuum.
This divergence can be regularized with an ultraviolet (UV) cutoff $\Lambda$ by subtracting the divergent part as
\begin{equation}
\Delta(x; \mu; \Lambda) \coloneqq \Delta(x; \mu) - \Delta(x; \Lambda) = \frac{1}{2\pi} \ln\frac{\Lambda}{\mu} + \mathcal{O}(x^2),
\end{equation}
leading to the well-known formula~\cite{Coleman:1974bu}
\begin{equation}
\mathcal{N}_{\mu} \exp(i \beta \phi(x)) = \qty(\Lambda / \mu)^{\beta^2 / 4\pi} \exp(i \beta\phi(x)). \label{eq:Coleman}
\end{equation}
The re-normal ordering formula~\cite{Coleman:1974bu}
\begin{equation}
\mathcal{N}_{\mu} \exp(i \beta \phi(x)) = \qty(\mu^\prime / \mu)^{\beta^2 / 4\pi} \mathcal{N}_{\mu^\prime} \exp(i \beta\phi(x)) \label{eq:renormal}
\end{equation}
is also obtained from Eq.~(\ref{eq:Coleman}).

In the infinitely large lattice system, the Feynman propagator 
\begin{align}
\Delta\qty(x; \mu; \frac{1}{a}) &= \int^{\pi}_{-\pi} \frac{d^2 k}{\qty(2 \pi)^2} e^{i k x} \Delta(k, ag), \\
\Delta(k, ag) &= \qty(4 \sum_{\mu} \sin^2\qty(\frac{k_\mu}{2}) + \qty(a\mu)^2)^{-1} \label{eq:Prop_mom}
\end{align}
is naturally regularized with the lattice spacing $a$.
By substituting it into Eq.~(\ref{eq:NO}), the lattice counterpart of Eq.~(\ref{eq:Coleman}) can be obtained~\cite{Bender:1984qg}:
\begin{align}
\mathcal{N}_{g / \sqrt{\pi}} \exp(i \beta \phi_x) &= \mathcal{O}(1 / ag)^{\beta^2 / 4\pi} \exp(i \beta \phi_x), \label{eq:Coleman_latt} \\
\mathcal{O}(1 / ag) &\coloneqq \exp{2 \pi \Delta\qty(0; \frac{g}{\sqrt{\pi}}; \frac{1}{a})}. \label{eq:UVdivFactor}
\end{align}
The factor $\mathcal{O}(1 / ag)$ defined here appears frequently in the following, and we call it the UV divergent factor in this paper since it is divergent in the continuum limit $ag \to 0$.
Table~\ref{table:uvdivfactor_theory} shows the actual values of the UV divergent factor (\ref{eq:UVdivFactor}) at various lattice spacings.
\begin{table}[!h]
\centering
\begin{tabular}{ccc}
\hline
$ag$ & $\mathcal{O}(1/ ag)$ & $ag \mathcal{O}(1 / ag)$ \\
\hline
2.8   & 2.962097... & 8.293871... \\
0.4   & 24.63885... & 9.855540... \\
0.2   & 49.86135... & 9.972271... \\
0.1   & 100.1014... & 10.01014... \\
0.025 & 401.0057... & 10.02514... \\
0.01  & 1002.625... & 10.02625... \\
0.001 & 10026.50... & 10.02650... \\
\hline
\end{tabular}
\caption{The UV divergent factors $\mathcal{O}(1 / ag)$ (\ref{eq:UVdivFactor}) at various lattice spacings.}
\label{table:uvdivfactor_theory}
\end{table}
We find the UV divergent factor behaves as $\mathcal{O}\qty(1 / ag) \simeq 10 / ag$ at $ag \ll 1$.

We can now define the lattice counterpart of the Hamiltonian~(\ref{eq:continuum_hamiltonian}) without using the normal ordering prescription
\begin{equation}
a H = \sum_{x = 0}^{L_x - 1} \frac{1}{2} \qty(a \pi_{x})^2 + \frac{1}{2} \qty( \partial_x \phi_{x} )^2 + 
\frac{\qty(ag)^2}{2\pi} \qty(\phi_{x} + \frac{\theta}{2 \sqrt{\pi}})^2
- \frac{e^{\gamma}}{2 \pi^{3/2}} \frac{m}{g} \qty(ag)^2 \mathcal{O}(1/ag) \cos(2 \sqrt{\pi} \phi_{x}), \label{eq:lattice_hamiltonian}
\end{equation}
where $\partial_x$ denotes the forward derivative $\partial_x f_x \coloneqq f_{x + 1} - f_x$.
Since the continuum model~(\ref{eq:continuum_hamiltonian}) is formulated on an infinite line,
the spatial length $L_x ag$ should be infinitely large in principle.
However, in practical numerical simulations, $L_x$ must be finite and some boundary condition must be specified.
In this paper, we impose the periodic boundary condition to preserve the translational symmetry and expect that the boundary condition becomes irrelevant in the large spatial length limit $L_x ag \to \infty$.

The thermal expectation value of an observable $O(\phi)$ at temperature $T / g = \qty(L_{\tau} ag)^{-1}$ can be expressed using the path-integral~\cite{Matsubara:1955ws}
\begin{subequations}
\begin{align}
\expval{O(\phi)} &= \left. \tr O(\phi) e^{-H / T} \middle/ \tr e^{-H / T} \right. \\
&= \left. \int D \phi \, O(\phi) e^{- S_{E}} \middle/ \int D \phi \, e^{-S_{E}}, \right. \label{eq:thermalexpvalue}
\end{align}
\end{subequations}
where $S_E$ is the lattice Euclidean action of the bosonized Schwinger model
\begin{multline}
S_E = \sum_{\tau = 0}^{L_\tau - 1} \sum_{x = 0}^{L_x - 1} \frac{1}{2} \qty( \partial_\tau \phi_{x, \tau})^2 + \frac{1}{2} \qty( \partial_x \phi_{x, \tau} )^2 +  \frac{\qty(ag)^2}{2\pi} \qty(\phi_{x, \tau} + \frac{\theta}{2\sqrt{\pi}})^2 \\
- \frac{e^{\gamma}}{2 \pi^{3/2}} \frac{m}{g} \qty(ag)^2 \mathcal{O}(1/ag) \cos(2 \sqrt{\pi} \phi_{x, \tau}). \label{eq:action}
\end{multline}
Here the periodic boundary condition must be imposed for the imaginary time direction since the scalar field is bosonic.
The lattice Euclidean action~(\ref{eq:action}) is obviously real and bounded below even at $\theta \neq 0$, meaning no sign problem.
The Monte Carlo configurations can be easily generated by combined use of the heat-bath algorithm and the rejection sampling, just like the case of $\mathrm{SU}(2)$ Yang--Mills theory~\cite{Creutz:1980zw}.

\section{Verification of the lattice formulation} \label{sec:vertification}
In this section, we verify the lattice bosonized Schwinger model~(\ref{eq:action}) by reproducing analytical and numerical results in the literature.
As an observable, we focus on the chiral condensate
\begin{subequations}
\begin{align}
\overline{\psi}\psi &= -\frac{e^\gamma}{2 \pi^{3/2}} g \mathcal{N}_{g / \sqrt{\pi}} \cos(2 \sqrt{\pi} \phi) \\
&= -\frac{e^\gamma}{2 \pi^{3/2}} g \mathcal{O}(1 / ag) \cos(2 \sqrt{\pi} \phi),
\end{align}
\end{subequations}
because it is directly related to the nontrivial normal ordering and is suitable for the check of the present lattice formulation.

\subsection{Analyical expression for the chiral condensate at $m = 0$} 
We first derive the analytical expression for the chiral condensate at $m = 0$
\begin{equation}
\expval{\overline{\psi}\psi}_{\mathrm{lat.}} = -\frac{e^\gamma}{2 \pi^{3/2}} g \mathcal{O}(1 / ag) \expval{\cos(2 \sqrt{\pi} \phi)}_{m = 0}.
\end{equation}
Using Wick's theorem, the thermal expectation value of $\exp(i\beta\phi)$ is analytically obtained as
\begin{equation}
\expval{\exp(i \beta \phi)}_{m = 0} = \exp{-\frac{\beta^2}{2} \Delta\qty(0; \frac{g}{\sqrt{\pi}}; \frac{1}{a})_{L_x, L_\tau}},
\end{equation}
where $\Delta(0; g / \sqrt{\pi}; 1/a)_{L_x, L_\tau}$ is the Feynman propagator in the lattice system with $L_x \times L_\tau$ sites in the periodic boundary conditions, and its explicit form reads
\begin{equation}
\Delta\qty(x; \frac{g}{\sqrt{\pi}}; \frac{1}{a})_{L_x, L_\tau} = \frac{1}{L_x L_\tau} \sum_{k_x = 2\pi / L_x, 4\pi / L_x, \dots}^{2\pi} \sum_{k_\tau = 2\pi / L_\tau, 4\pi / L_\tau, \dots}^{2\pi} e^{i k x} \Delta(k, ag).
\end{equation}
Hence, we obtain 
\begin{equation}
\expval{\overline{\psi}\psi}_{\mathrm{lat.}} = -\frac{e^\gamma}{2 \pi^{3/2}} g
\exp[-2\pi\qty{\Delta\qty(0; \frac{g}{\sqrt{\pi}}; \frac{1}{a})_{L_x, L_\tau} - \Delta\qty(0; \frac{g}{\sqrt{\pi}}; \frac{1}{a})}]. \label{eq:analytical_cc}
\end{equation}
At zero temperature in the large spatial length limit ($L_x ag, L_\tau ag \to \infty)$,
the argument of the exponential becomes zero, 
and the analytically exact chiral condensate
\begin{equation}
\expval{\overline{\psi} \psi}_{\mathrm{cont.}} = -\frac{e^\gamma}{2\pi^{3/2}} g 
\end{equation}
is reproduced. 
In the large spatial length and continuum limits ($L_x ag, 1 / ag \to \infty$), the analytical chiral condensate~(\ref{eq:analytical_cc}) should converge to that obtained by directly evaluating the fermionic path-integral in the continuum formulation~\cite{Sachs:1991en}
\footnote{This expression first appeared in Ref.~\cite{Hetrick:1988yg}, 
in which the chiral condensate in the massless Schwinger model on a circle ($0 \le x < L$) was analytically obtained using bosonization in the Hamiltonian formalism. 
Because of the equivalence between space and time in (1 + 1)-dimensional Euclidean space-time, the resultant chiral condensate is equivalent to that at temperature $T = 1 / L$ in the large spatial length limit.}
\begin{align}
\expval{\overline{\psi} \psi}_{\mathrm{cont.}} &= -\frac{e^\gamma}{2\pi^{3/2}} g \exp{2 I\qty(\frac{g}{\sqrt{\pi} T})}, \label{eq:analytic_SW} \\
I(x) &\coloneqq \int_0^\infty dt \, \qty(1 - e^{x \cosh t})^{-1}.
\end{align}

\begin{figure}[htb]
\centering
\includegraphics[width=1.0\textwidth]{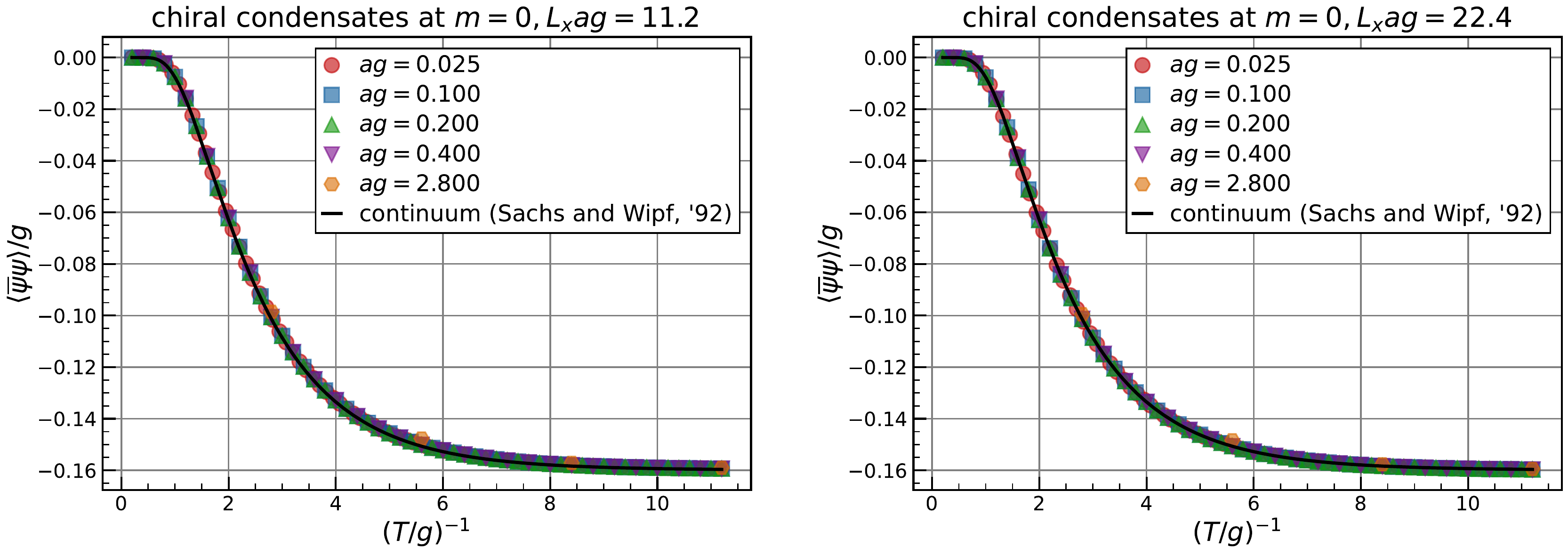}

\includegraphics[width=1.0\textwidth]{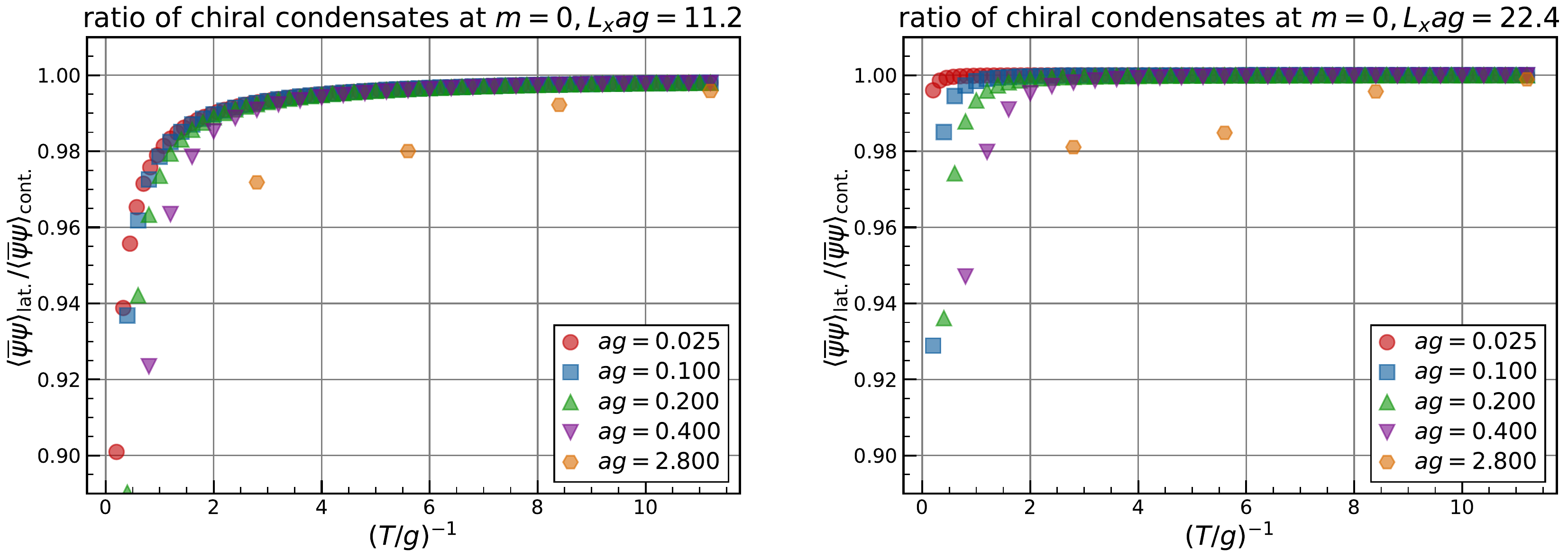}
\caption{
(Upper half) The two analytical chiral condensates (\ref{eq:analytical_cc}, \ref{eq:analytic_SW}) as a function of temperature at $L_x ag = 11.2$ (left) and $L_x ag = 22.4$ (right). 
(Lower half) Ratio of the two analytical chiral condensates at the same parameters.
\label{fig:masslessChiralCondensate}
}
\end{figure}

We can investigate the finite lattice spacing and spatial length effects at $m = 0$ by comparing Eqs.~(\ref{eq:analytical_cc}) and (\ref{eq:analytic_SW}).
The upper half of Fig.~\ref{fig:masslessChiralCondensate} shows the two analytical chiral condensates~(\ref{eq:analytical_cc}, \ref{eq:analytic_SW}) at various lattice spacings $ag$ and two spatial lengths $L_x ag = 11.2, 22.4$.
Remarkably, we find that the lattice chiral condensate~(\ref{eq:analytical_cc}) appears to agree with the continuum one~(\ref{eq:analytic_SW}) even at a very large lattice spacing $ag = 2.8$.
For a more detailed investigation, we show the ratio of the two analytical chiral condensates (\ref{eq:analytical_cc}, \ref{eq:analytic_SW}) at the same parameters in the lower half of Fig.~\ref{fig:masslessChiralCondensate}.
We find that the $ag$ dependence is almost negligible for $ag \le 0.2$. 
At $L_x ag = 11.2$, the spatial length is not satisfactory large, 
and some discrepancies can be seen.
These discrepancies are almost absent at $L_x ag = 22.4$.
Note that the small discrepancies at high temperatures $(T / g)^{-1} \lesssim 1$ should not be taken seriously because the chiral condensate is almost zero at these temperatures.
We conclude that both continuum and large spatial length limits are reliably taken for $ag \lesssim 0.2, L_x ag = 22.4$ at $m = 0$.
In the following numerical simulations at $m \neq 0$ in this section, we very conservatively use the lattice of $ag = 0.025, L_x = 896$ and generate $N_\mathrm{conf} = 10^6$ Monte Carlo configurations for each measurement, unless otherwise mentioned.

\subsection{Chiral condensate at $m \neq 0$}
We next calculate the chiral condensate at $m \neq 0, \theta = 0$.
While there exists no analytically exact result,
the chiral condensates at both zero and nonzero temperatures have been extensively studied using the tensor network method~\cite{Buyens:2014pga, Banuls:2016lkq, Buyens:2016ecr}.
We compare our results with theirs and check the lattice formulation.
We stress that this serves as another nontrivial check since we are dealing with the normal ordering dynamically in this case.
In this subsection, we remove the logarithmic divergence in the chiral condensate at $m \neq 0$ by subtracting the free chiral condensate at (almost) zero temperature following Refs.~\cite{Buyens:2014pga, Banuls:2016lkq, Buyens:2016ecr}.
We use the jackknife method to estimate statistical errors.

The chiral condensates at $T / g = \qty(448 \times 0.025)^{-1}$ obtained in this work and the most recent results by the tensor network method at zero temperature~\cite{Banuls:2016lkq} are summarized in Table~\ref{table:chiralcondensate}.
Our numerical results match theirs with approximately one percent accuracy.
It is notable that our results are obtained with no continuum nor infinite spatial length extrapolation in contrast to the tensor network calculations, although the errors are far larger. 
If one aims at the precision of Ref.~\cite{Banuls:2016lkq} using the current lattice parameters, 
around $N_{\mathrm{conf}} = 10^{10}\text{--}10^{12}$ configurations are required, which is not practically feasible.

\begin{table}[!h]
\centering
\begin{tabular}{cccc}
\hline
$m/g$ & This work & Ref.~\cite{Banuls:2016lkq} & \hspace{-0.6cm} This work / Ref.~\cite{Banuls:2016lkq} \\
\hline
0.0625 & 0.11506(91) & 0.1139657(8)  & 1.0096(80) \\
0.125  & 0.09249(66) & 0.0920205(5)  & 1.0051(72) \\
0.25   & 0.06629(62) & 0.0666457(3)  & 0.9947(93) \\
0.5    & 0.04207(37) & 0.0423492(20) & 0.9935(87) \\
1      & 0.02385(22) & 0.0238535(28) & 0.9997(93) \\
\hline
\end{tabular}
\caption{Absolute values of the chiral condensates at $T / g = \qty(448 \times 0.025)^{-1}$ obtained in this work, compared with the tensor network results at zero temperature~\cite{Banuls:2016lkq}.}
\label{table:chiralcondensate}
\end{table}

\begin{figure}[htb]
\centering
\includegraphics[width=0.45\textwidth]{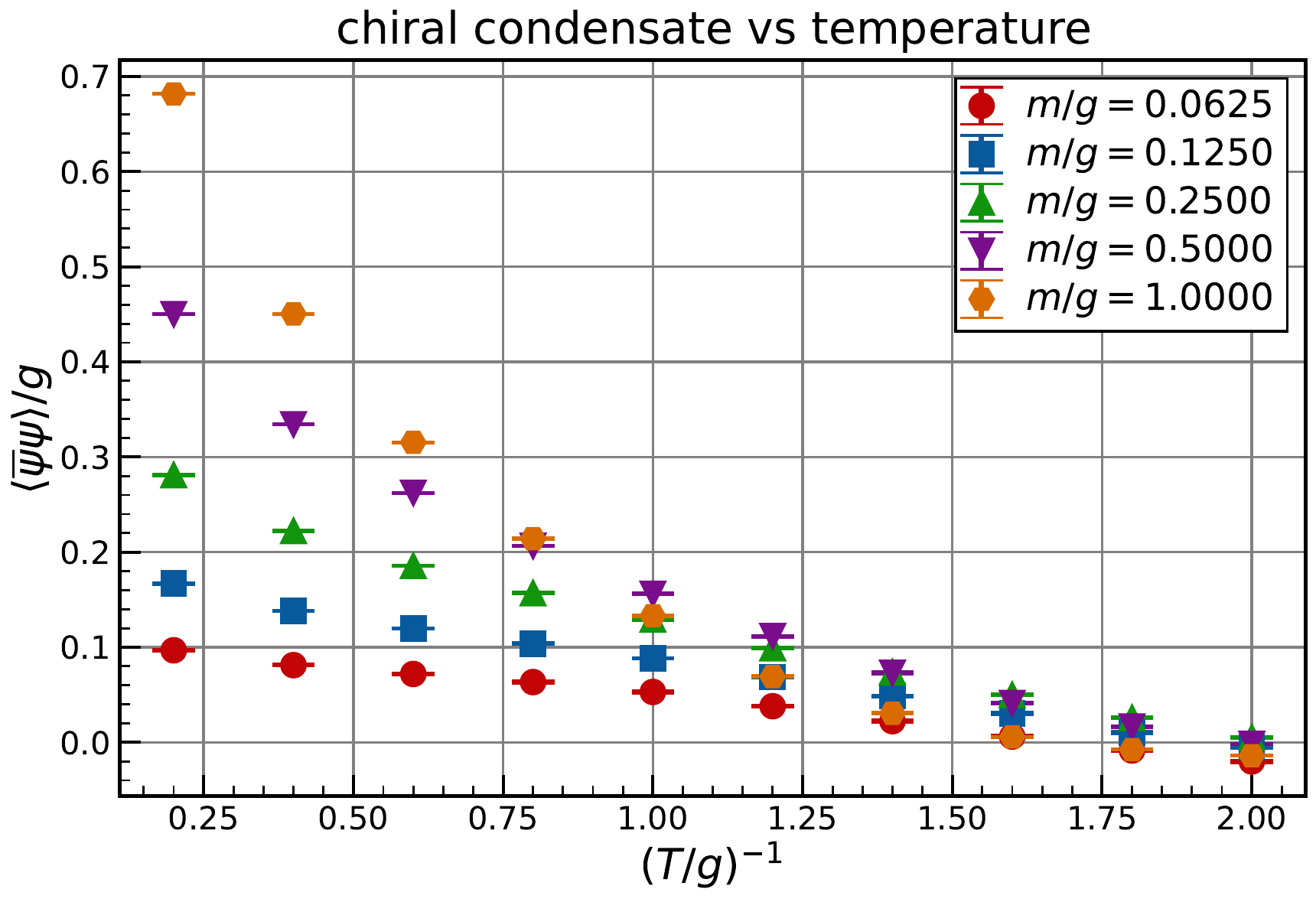}
\caption{
Temperature dependence of the chiral condensate at $m / g = 0.0625, 0.125, 0.25, 0.5, 1$. 
\label{fig:temperature_dependence}
}
\end{figure}
The present method is advantageous at finite temperatures.
Figure~\ref{fig:temperature_dependence} shows the temperature dependence of the chiral condensate at $m / g = 0.0625, 0.125, 0.25, 0.5, 1$.
Our numerical results are seemingly consistent with the tensor network results~\cite{Banuls:2016lkq, Buyens:2016ecr} (see Fig.~4 in Ref.~\cite{Buyens:2016ecr}), and we achieve better precision at high temperatures.
Those results provide further evidence that the lattice formulation of the bosonized Schwinger model is valid.

\subsection{Finite $\theta$}
\begin{figure}[htb]
\centering
\includegraphics[width=0.45\textwidth]{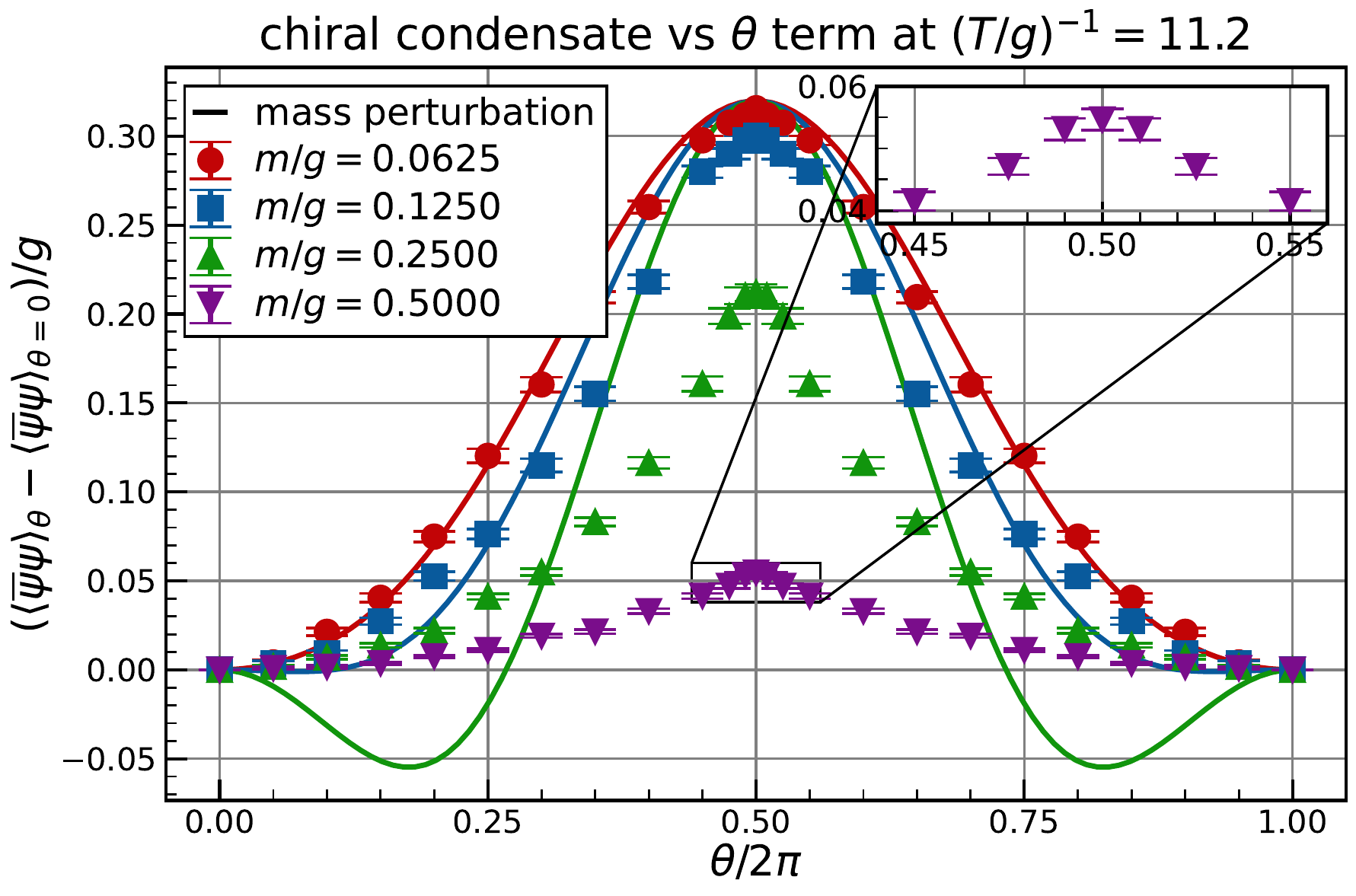}
\caption{
$\theta$ dependence of the chiral condensate at $m / g = 0.0625, 0.125, 0.25, 0.5, T / g = \qty(448 \times 0.025)^{-1}$. 
The chiral condensates at $m / g = 0.0625, 0.125, 0.25$ are compared with the leading-order mass perturbation~(\ref{eq:mass_perturbation}) (solid line).
\label{fig:theta_dependence}
}
\end{figure}

After verifying the formulation, we next study the Schwinger model with a $\theta$ term, 
which is inaccessible by the standard Monte Carlo simulation based on the conventional lattice formulation due to the sign problem.
In the present method, no difficulty exists.
Figure~\ref{fig:theta_dependence} shows the $\theta$ dependences of the chiral condensates
\begin{equation}
\expval{\overline{\psi}\psi}_\theta = -\frac{e^{\gamma}}{2 \pi^{3/2}} g \mathcal{O}(1 / ag) \expval{\cos(2 \sqrt{\pi}\phi)}_\theta
\end{equation}
at $m / g = 0.0625, 0.125, 0.25, 0.5$.
Here Monte Carlo configurations only at $\theta \le \pi$ are generated, and data points at $\theta > \pi$ are obtained using the line symmetry at $\theta = \pi$.
The statistical errors are all smaller than the symbols, even though the chiral condensates at nonzero $\theta$ are evaluated using $N_{\mathrm{conf}} = 10^5$ configurations.
The chiral condensates at $m / g = 0.0625, 0.125, 0.25$ are compared with the leading-order mass perturbation~\cite{Adam:1997wt, Adam:1998tw}
\begin{equation}
\expval{\overline{\psi} \psi}_{\theta} - \expval{\overline{\psi} \psi}_{\theta = 0} = \frac{e^\gamma}{2 \pi^{3/2}} g \qty(1 - \cos\theta) - 0.358 m \qty(1 - \cos 2\theta).
\label{eq:mass_perturbation}
\end{equation}
The mass perturbation theory works well at $m / g = 0.0625, 0.125$, whereas sizable deviations appear at $m / g = 0.25$.
These behaviors are consistent with the tensor network results at zero temperature~\cite{Buyens:2017crb, Funcke:2019zna}, 
although we are calculating the chiral condensates at very low yet not zero temperature $T / g = (448 \times 0.025)^{-1}$.

A cusp-like behavior is observed at $\theta = \pi, m / g = 0.5$ in Fig.~\ref{fig:theta_dependence}, 
which might suggest the spontaneous CP symmetry breaking.
It is well known that the spontaneous CP symmetry breaking occurs at zero temperature for sufficiently large fermion masses $m / g \gtrsim 0.33$~\cite{Coleman:1976uz, Hamer:1982mx, Ranft:1982bi, Byrnes:2002nv, Shimizu:2014fsa, Buyens:2017crb, Azcoiti:2017mxl, Thompson:2021eze}.
The analogy with the quantum Ising chain and a tensor network study~\cite{Buyens:2016ecr} suggest that the CP symmetry is restored at any nonzero temperature.
To give a definitive answer to this issue, a very careful finite-size scaling analysis is required.
Although such a study would be meaningful, 
we leave it as a future study and turn our attention to confining properties in the Schwinger model at finite temperature and $\theta$.

\section{Confinement at finite temperature and $\theta$} \label{sec:confinement}
We finally investigate confinement/deconfinement properties in the Schwinger model at finite temperature and $\theta$.
For this purpose, we calculate the string tension in the $(T, \theta)$ plane.

Let us explain our method to calculate the string tension.
Because the $\theta$ angle is physically interpreted as the background electric field $E_{\mathrm{ex}} = (\theta / 2\pi)g$~\cite{Coleman:1976uz}, 
the inclusion of the two static probe charges $q_p g, -q_p g$ separated infinity can be described through a modification of the $\theta$ angle:
\begin{equation}
\theta \to 2 \pi q_p + \theta. 
\end{equation}
The string tension (coefficient of the linear term in the static potential) between two static probe charges $q_p g, -q_p g$ can be obtained from the difference in free energy densities
\begin{equation}
\sigma(q_p, \theta) = f(2\pi q_p + \theta) - f(\theta) = \frac{-1}{L_x L_\tau a^2} \ln \frac{Z(2\pi q_p + \theta)}{Z(\theta)},
\end{equation}
where $Z(\theta)$ is the partition function 
\begin{equation}
Z(\theta) = \tr e^{-H(\theta) / T} \propto \int D\phi\, e^{-S_E(\theta)}.
\end{equation}
The computation of the free energy density itself is difficult by the Monte Carlo method since it cannot be expressed as an expectation value. However, the difference can be:
\begin{equation}
\sigma(q_p, \theta) 
= \frac{-1}{L_x L_\tau a^2} \ln \expval{ \exp[-\frac{\qty(ag)^2}{\sqrt{\pi}} \sum_{x, \tau} 
q_p \qty(\phi_{x, \tau} + \frac{\theta + \pi q_p}{2\sqrt{\pi}})
]
}_{\theta}. \label{eq:direct_formula}
\end{equation}
In practical numerical simulations, direct evaluation using Eq.~(\ref{eq:direct_formula}) leads to large statistical and systematic errors at large $q_p$.
To avoid this problem, we consider to decompose the string tension at
\begin{equation}
q_p = N \delta q_p, \quad \theta = M \delta \theta, \quad N, M \in \mathbb{Z},
\end{equation}
where $\delta q_p, \delta \theta$ are some small step widths.
When $\delta q_p = \delta \theta / 2\pi$,
the free energy density can be expressed as
\begin{equation}
f(2\pi q_p + \theta) = f((N + M) \delta \theta) \eqqcolon f_{N + M}.
\end{equation}
Using this notation, the string tension can be decomposed as follows
\begin{subequations}
\begin{align}
\sigma(N \delta q_p, M \delta\theta) &= f_{N + M} - f_{M} \\
&= \sum_{i = 0}^{N - 1} f_{i + M + 1} - f_{i + M} \\
&= \frac{-1}{L_x L_\tau a^2} \sum_{i = 0}^{N - 1} \ln \expval{\exp[-\frac{\qty(ag)^2}{\sqrt{\pi}} \sum_{x, \tau} \delta q_p \qty( \phi_{x, \tau} + \frac{\theta + \pi \delta q_p}{2\sqrt{\pi}})]}_{\theta = \qty(i + M) \delta \theta}. \label{eq:indirect_formula}
\end{align}
\end{subequations}
By using Eq.~(\ref{eq:indirect_formula}) instead of Eq.~(\ref{eq:direct_formula}), we can greatly mitigate the large statistical and systematic errors at large $q_p$ since $q_p$ in Eq.~(\ref{eq:direct_formula}) is now replaced by $\delta q_p$.

\begin{figure}[htb]
\centering
\includegraphics[width=1.0\textwidth]{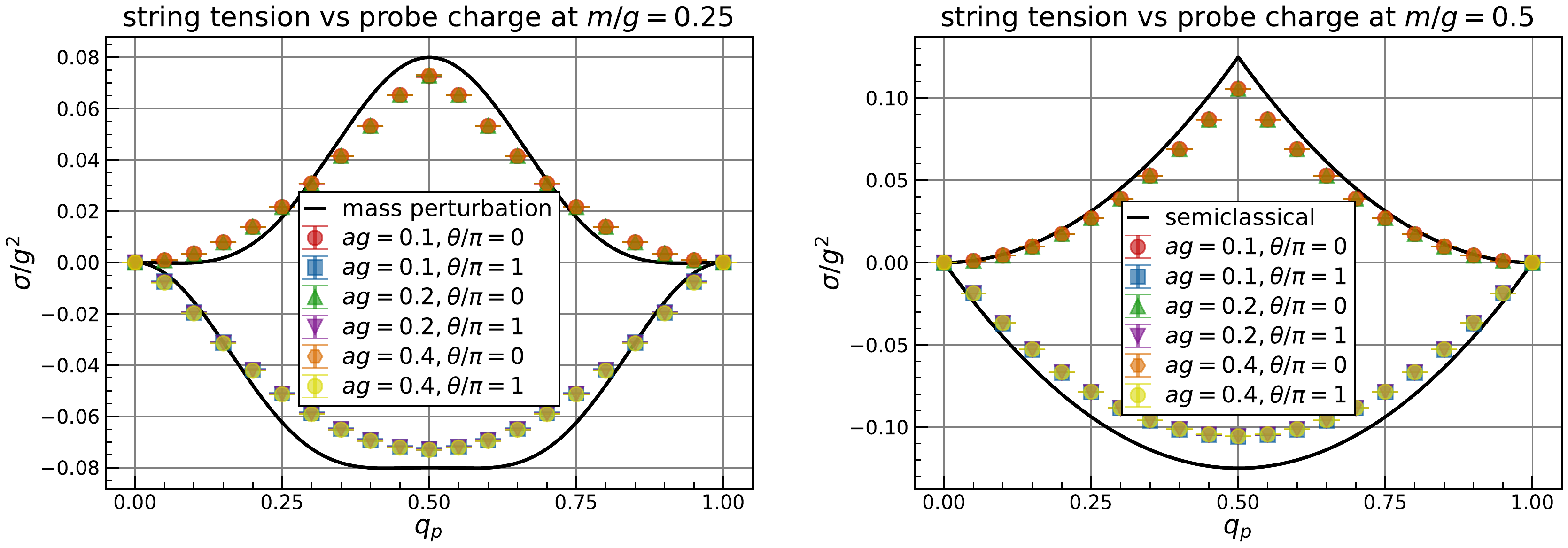}
\caption{
Probe charge dependence of the string tension for $\theta = 0, \pi$ at $m / g = 0.25$ (left) and $m / g = 0.5$ (right). 
The string tensions at $m / g = 0.25$ and $m / g = 0.5$ are compared with the mass perturbation~(\ref{eq:string_mass}) and the semiclassical estimates~(\ref{eq:semiclassical_string}), respectively.
The temperature and spatial length are held constant at $T / g = 11.2^{-1}$ and $L_x ag = 22.4$, respectively.
\label{fig:probe_charge_dep}
}
\end{figure}

Figure~\ref{fig:probe_charge_dep} shows the probe charge dependence of the string tension at $ag = 0.4, 0.2, 0.1$ and $m / g = 0.25, 0.5$.
The results are obtained from $N_{\mathrm{conf}} = 10^6$ configurations at $\theta / \pi$ ranging from $0.0$ to $1.0$ with a step width of $\delta \theta / \pi = 0.1$. 
The temperature and spatial length are held constant at $T / g = 11.2^{-1}$ and $L_x ag = 22.4$, respectively.
We find that the results at $ag = 0.4, 0.2, 0.1$ exhibit exceptional precision and agreement.
Motivated by this, we use the lattice of $ag = 0.2, L_x = 112$ in the following analysis.
\begin{figure}[htb]
\centering
\includegraphics[width=0.7\textwidth]{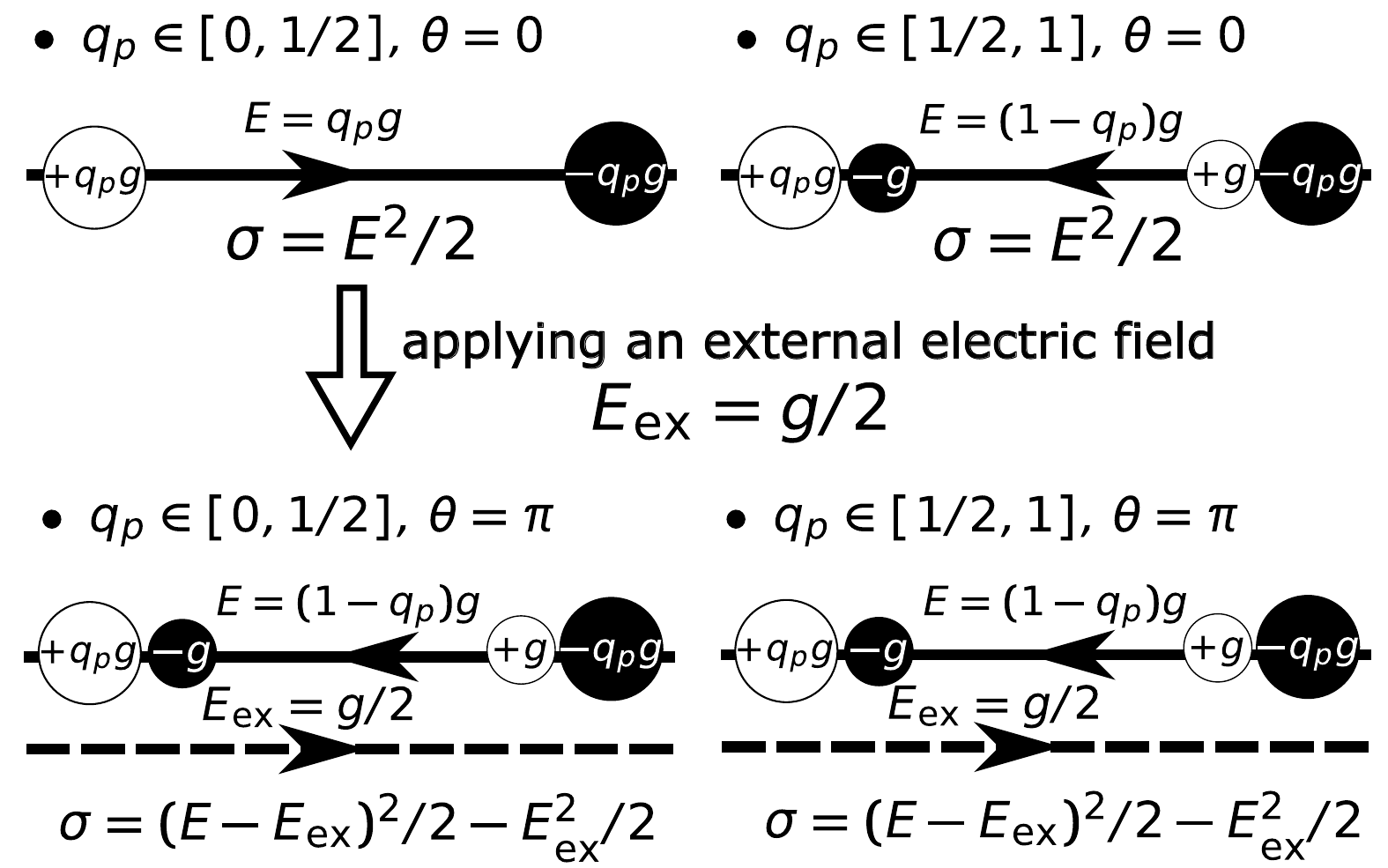}
\caption{
Semiclassical picture of the string tension between two static probe charges $q_p g, -q_p g$.
\label{fig:semiclassical}
}
\end{figure}

In Fig.~\ref{fig:probe_charge_dep}, we observe peculiar behaviors in the string tension: at $\theta = 0$, there is a peak in the string tension at $q_p = 0.5$. More interestingly, the string tension becomes negative for noninteger probe charges at $\theta = \pi$.
These behaviors can be well understood through semiclassical analysis of the string tension (see Fig.~\ref{fig:semiclassical}).
For $q_p \in \qty[0, 1/2], \theta = 0$, a constant electric field $E = q_p g$ appears between the two static probe charges at the classical level due to the Gauss law, as shown in the upper left of Fig.~\ref{fig:semiclassical}.
Consequently, the string tension increases quadratically as the probe charge $q_p$ increases.
When $q_p$ exceeds $1/2$, and the distance between the probe charges is sufficiently large, the vacuum produces a fermion-antifermion pair, reducing the total electric field by forming a two ``meson" system (upper right of Fig.~\ref{fig:semiclassical}).
Setting $\theta = \pi$, i.e., applying an external electric field $E_\mathrm{ex} = g/2$, to the two ``meson" system,
the external electric field works to decrease the total electric field, resulting in the negative string tension.
In the case of the single ``meson" system (upper left of Fig.~\ref{fig:semiclassical}), as the external electric field approaches $g / 2$, the vacuum would again produce a fermion-antifermion pair to decrease the total electric field by forming a two ``meson`` system at a certain value of $\theta$. 
Therefore, at $\theta = \pi$, the configuration is the same regardless of the probe charge, as shown in the lower of Fig.~\ref{fig:semiclassical}.
The resulting semiclassical estimate for the string tension is then given by
\begin{equation}
\sigma / g^2 = 
\begin{cases}
\frac{1}{2} q_p^2 , & q_p \in \qty[0, \frac{1}{2}], \, \theta = 0, \\
\frac{1}{2} \qty(1 - q_p)^2 , & q_p \in \qty[\frac{1}{2}, 1], \, \theta = 0, \\
-\frac{1}{2} q_p \qty(1 - q_p), & q_p \in \qty[0, 1], \, \theta = \pi.
\end{cases} \label{eq:semiclassical_string}
\end{equation}
In the right panel of Fig.~\ref{fig:probe_charge_dep}, we find that the semiclassical string tension~(\ref{eq:semiclassical_string}) successfully explains the qualitative behaviors of our numerical results.

While it is hard to give an intuitive explanation for the negative string tension at small fermion mass, the next-to-leading order mass perturbation~\cite{Adam:1996rd, Adam:1997wt} 
\begin{equation}
\sigma / g^2 = \frac{e^\gamma}{2\pi^{3/2}} \frac{m}{g} \qty(\cos(\theta) - \cos(2\pi q_p + \theta)) - 0.179 
 \qty(\frac{m}{g})^2 \qty(\cos(2\theta) - \cos(4\pi q_p + 2\theta)) \label{eq:string_mass}
\end{equation}
successfully explains the qualitative behaviors, as shown in the left panel of Fig.~\ref{fig:probe_charge_dep}.
We note that the string tensions at $\theta = 0$ at various masses, probe charges, and temperatures have been already obtained with high precision by the tensor network method~\cite{Buyens:2015tea, Buyens:2016ecr, Buyens:2017crb}.
However, the string tension at nonzero $\theta$ has not been investigated well so far. 
In the case of the charge-$3$ Schwinger model, the string tension between integer probe charges at nonzero $\theta$ was studied in Ref.~\cite{Honda:2021ovk} through quantum simulation on a classical simulator.
Negative string tension was observed at large $\theta$, 
although reliable continuum extrapolation was difficult due to a limited number of lattice sites ($N \le 25$) and slow convergence to the continuum limit.
In Refs.~\cite{Misumi:2019dwq, Honda:2021ovk}, the negative string tension between integer probe charges in the charge-$q$ Schwinger model, where $q$ is an integer larger than $1$, was explained in terms of the $\mathbb{Z}_q$ 1-form symmetry. 
Unfortunately, their argument can not be applied to the present case.
Nevertheless, our numerical results demonstrate that the negative string tension appears for noninteger probe charges at almost zero temperature in the standard Schwinger model.

After confirming the effectiveness of our method in calculating the string tension and giving an intuitive physical explanation,
we explain our simulation strategy to establish the string tension in the $(T, \theta)$ plane.
To cover almost the entire $(T, \theta)$ plane, we generate $N_{\mathrm{conf}} = 10^6$ Monte Carlo configurations at $L_\tau = 4, 6, 8, 10, 12, 16, 20, 24, 28, 40, 56$, with $\theta / \pi$ ranging from $0.0$ to $1.0$ with a step width of $\delta \theta / \pi = 0.1$.
By combining these configurations with the reweighing method, we achieve a very smooth surface in the $(T, \theta)$ plane.
For both $T$ and $\theta$ directions, we obtain ten data points between adjacent simulation points, each reweighted from the nearest simulation point.
This results in $12^2$ data points within the unit cell formed by four simulation points.

\begin{figure}[htb]
\centering
\includegraphics[width=0.45\textwidth]{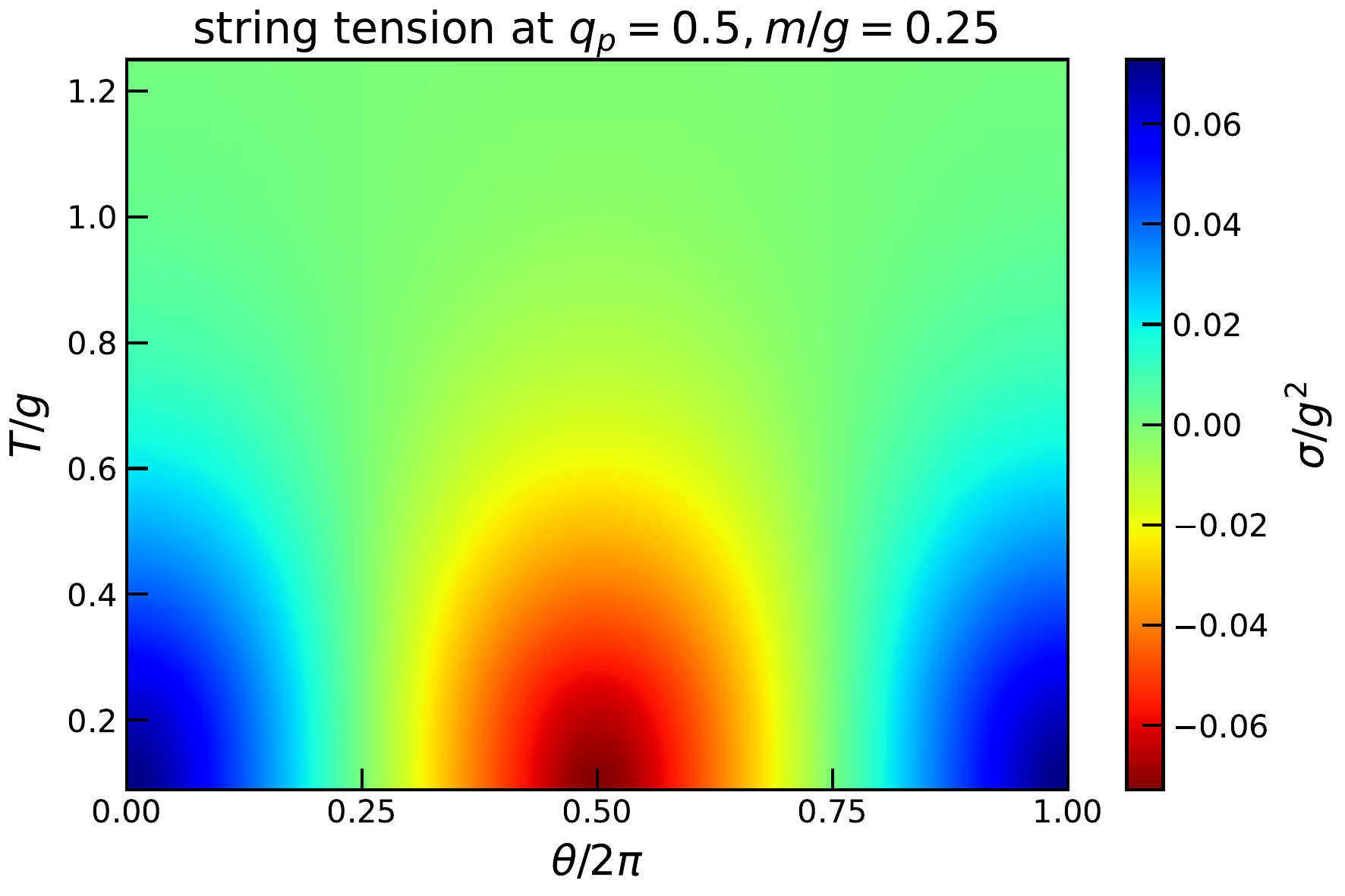}
\qquad
\includegraphics[width=0.45\textwidth]{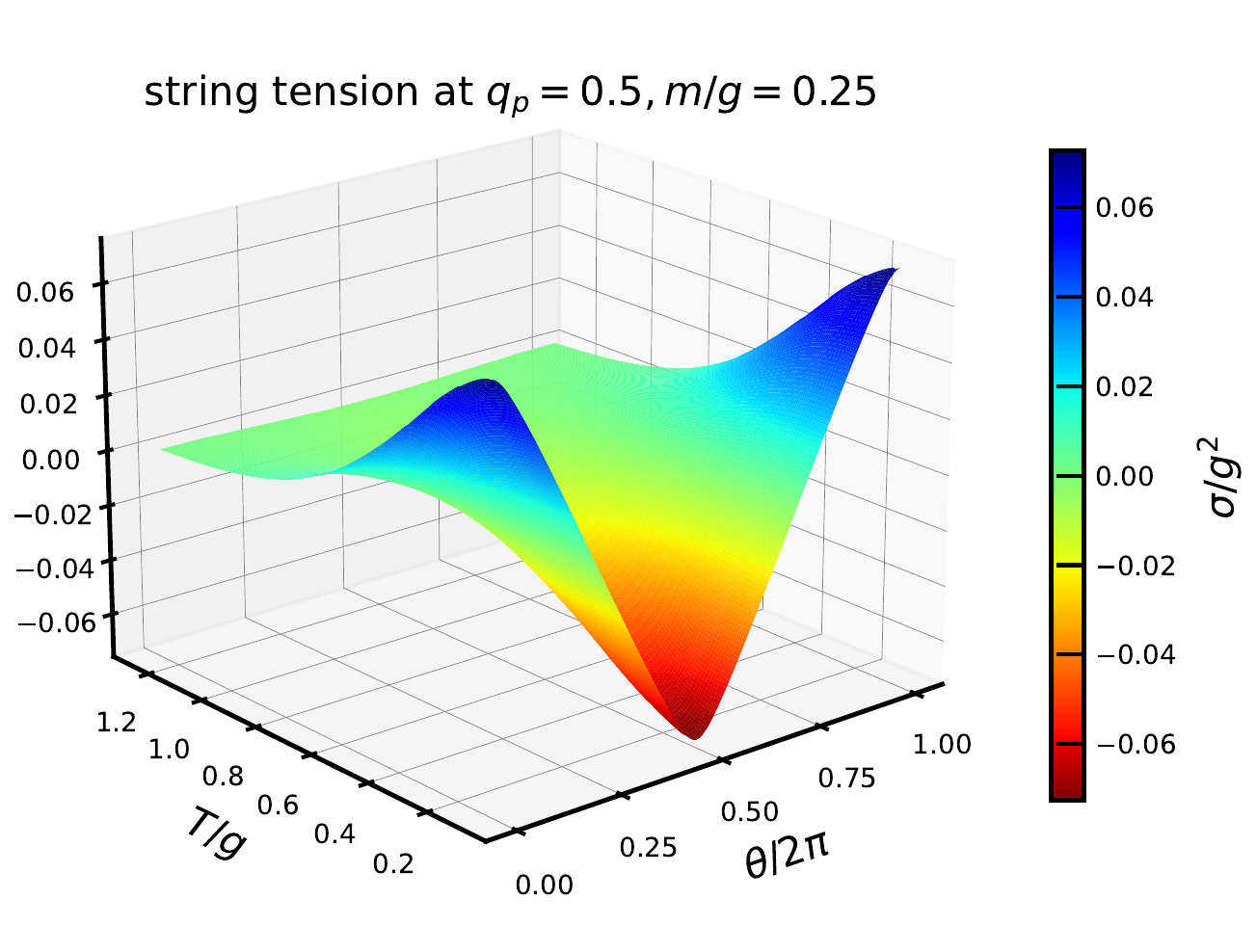}

\includegraphics[width=0.45\textwidth]{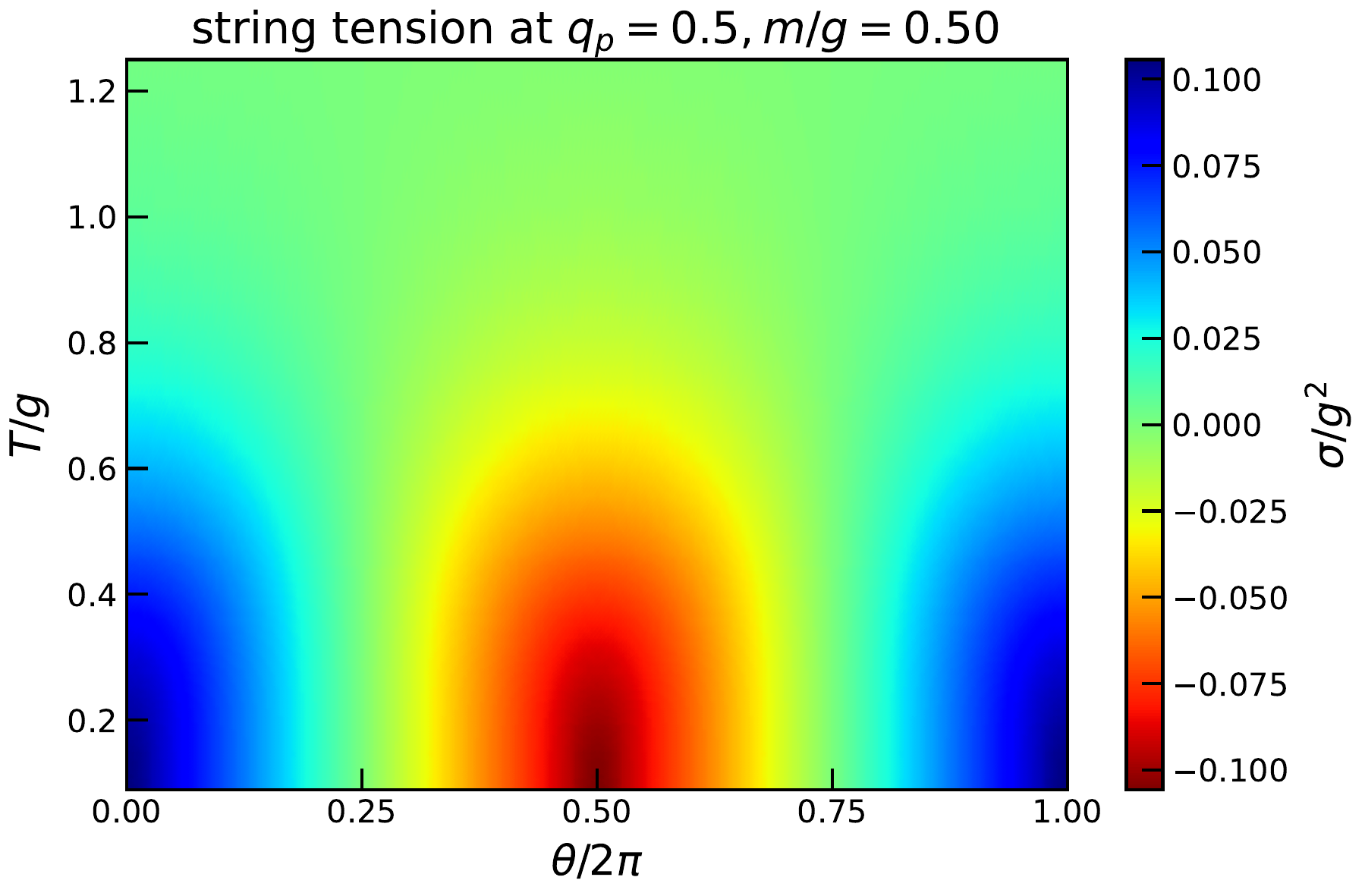}
\qquad
\includegraphics[width=0.45\textwidth]{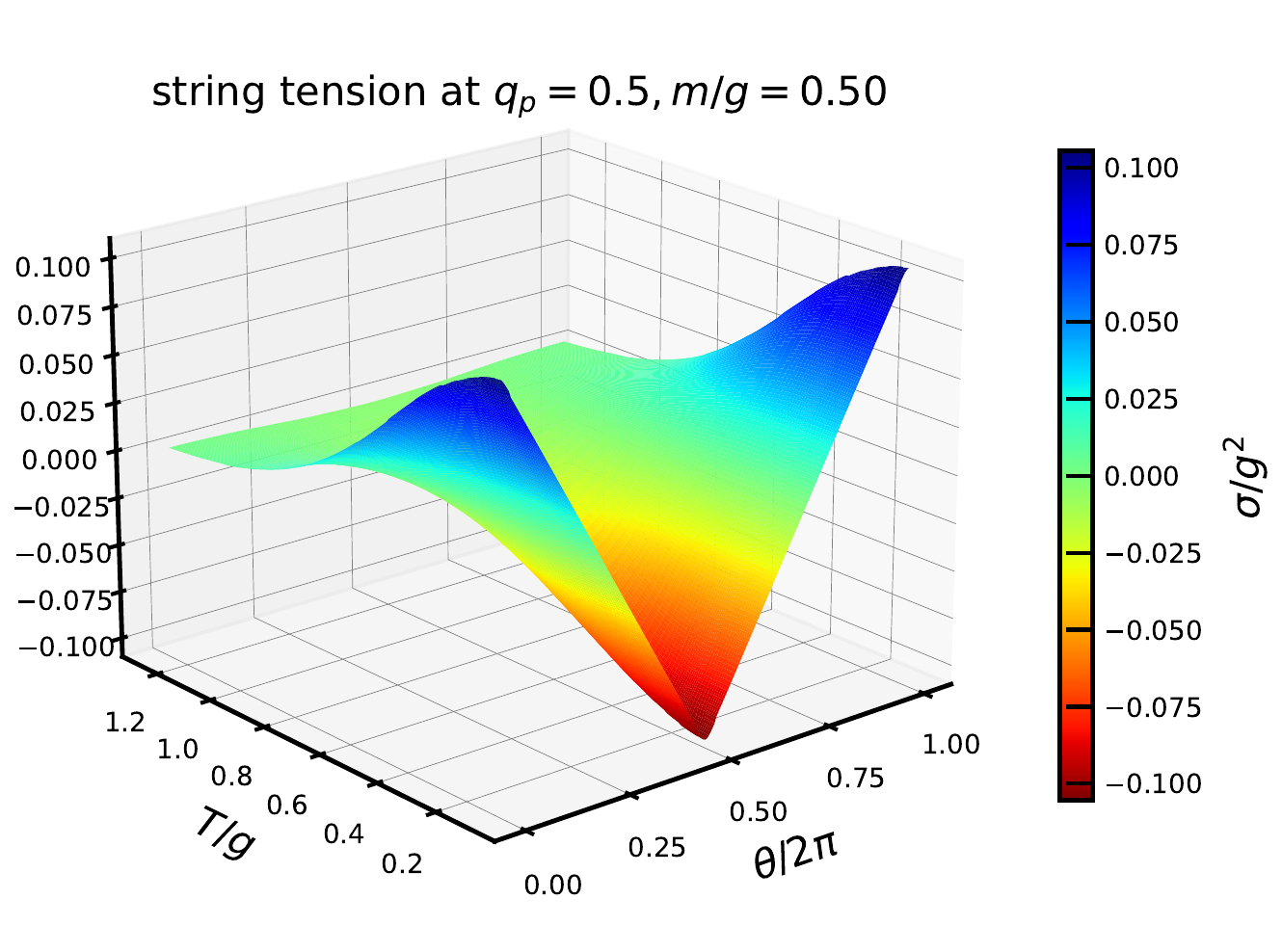}
\caption{
The string tension at $q_p = 0.5$ in the $(T, \theta)$ plane at $m / g = 0.25$ (upper half) and $m / g = 0.5$ (lower half).
\label{fig:3d_stringTension_0.5}
}
\end{figure}
\begin{figure}[htb]
\centering
\includegraphics[width=0.45\textwidth]{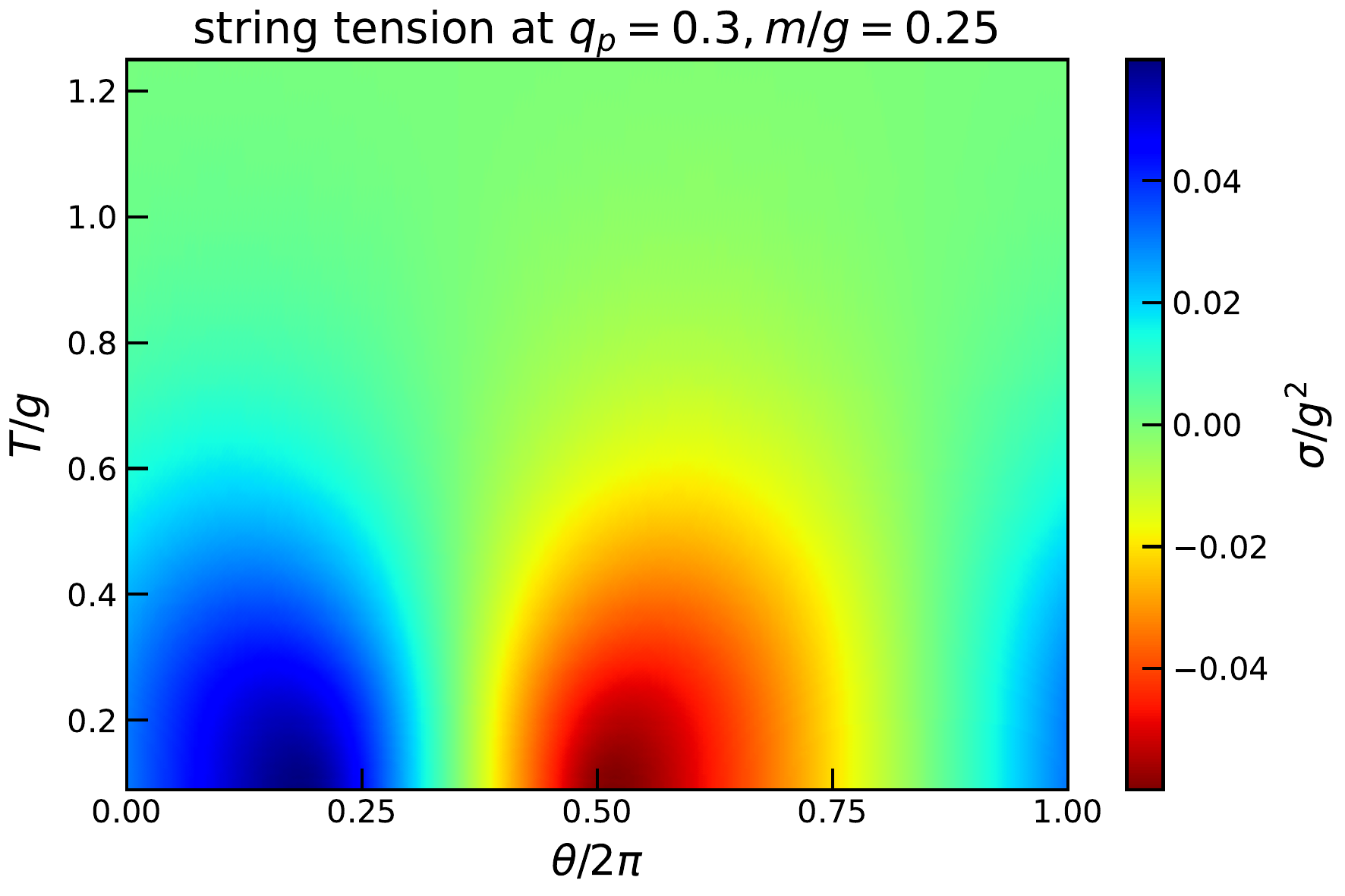}
\qquad
\includegraphics[width=0.45\textwidth]{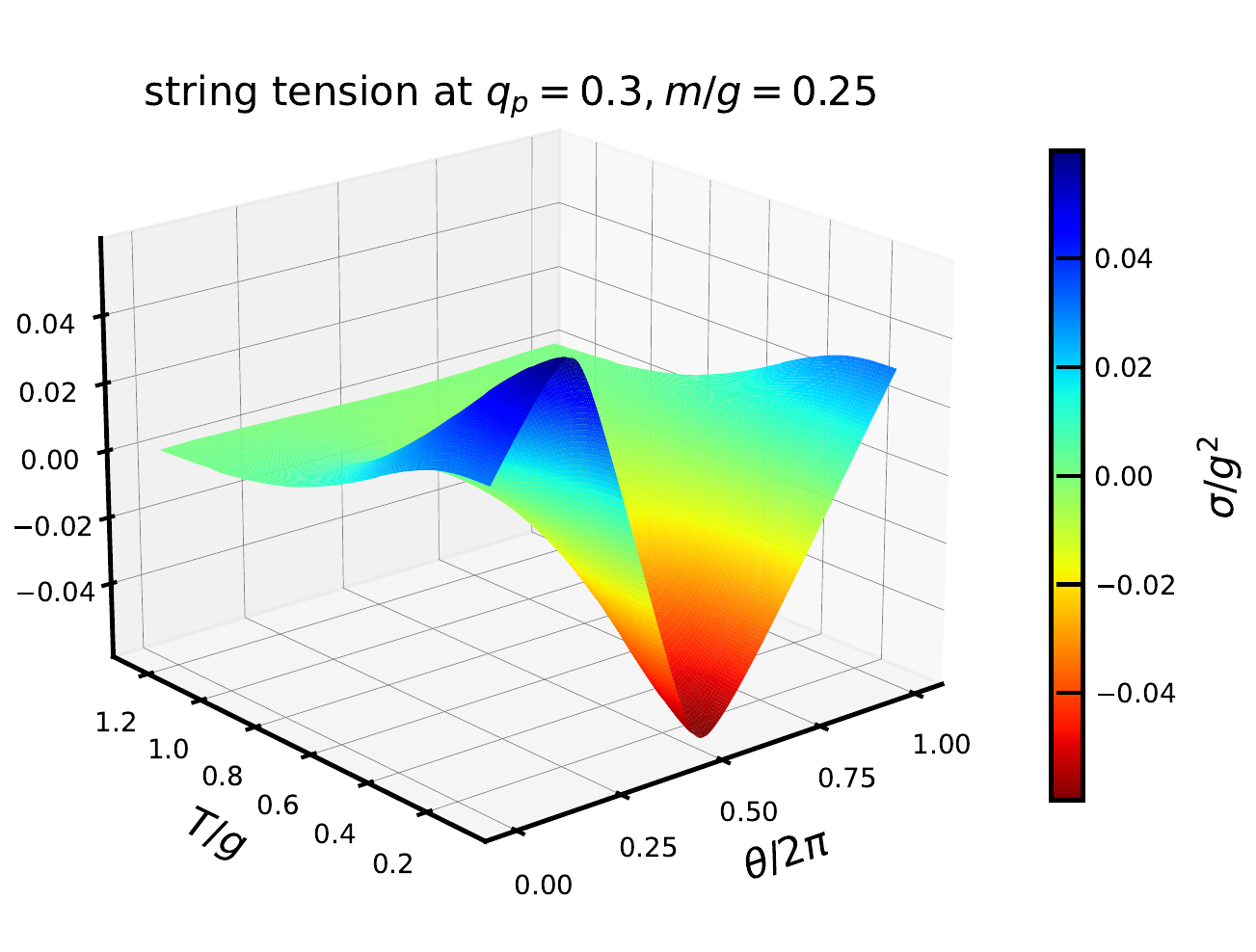}

\includegraphics[width=0.45\textwidth]{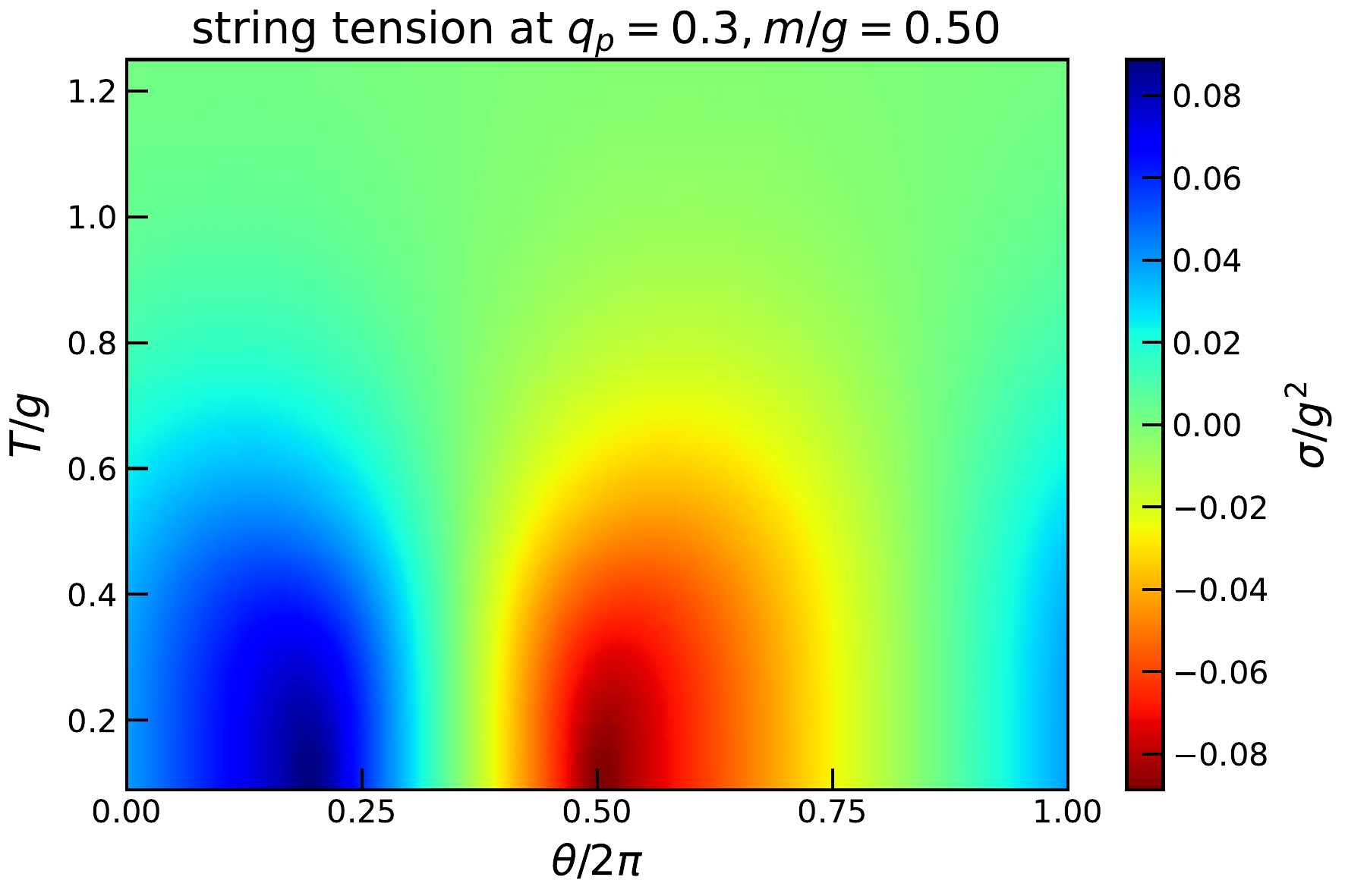}
\qquad
\includegraphics[width=0.45\textwidth]{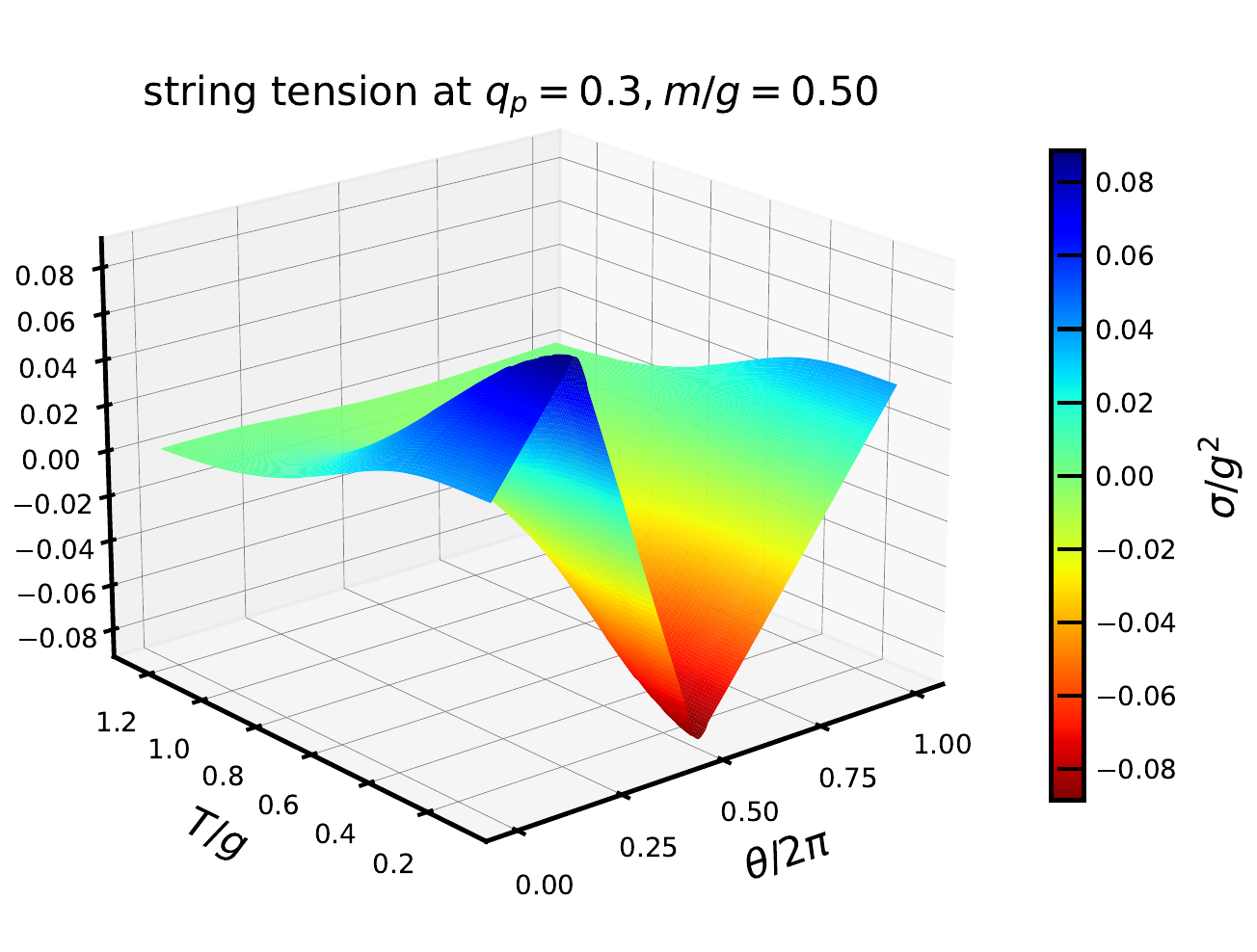}
\caption{
The string tension at $q_p = 0.3$ in the $(T, \theta)$ plane at $m / g = 0.25$ (upper half) and $m / g = 0.5$ (lower half).
\label{fig:3d_stringTension_0.3}
}
\end{figure}

Figure~\ref{fig:3d_stringTension_0.5} shows the string tension at $q_p = 0.5$ in the $(T, \theta)$ plane at $m / g = 0.25$ (upper half) and $m / g = 0.5$ (lower half).
At $q_p = 0.5$, one can easily show that 
\begin{equation}
\sigma(0.5, \theta + \pi) = - \sigma(0.5, \theta). \label{eq:0.5}
\end{equation}
For both $m / g = 0.25$ and $m / g = 0.5$, the string tension is positive around $\theta \simeq 0$ at low temperatures, indicating confinement.
The string tension diminishes as $\theta$ increases and becomes zero at $\theta / \pi = 0.5$.
With further increases in $\theta$, the string tension becomes negative and reaches its minimum at $\theta = \pi$.
The peak height is roughly proportional to the fermion mass $m / g$.
As temperature increases, the string tension gradually converges to zero at all $\theta$, indicating deconfinement.
In Fig.~\ref{fig:3d_stringTension_0.3}, we show similar plots but at $q_p = 0.3$, where simple constraint like Eq.~(\ref{eq:0.5}) does not exist.
Consequently, we observe shifts of the peak positions.
Nevertheless, the basic pattern remains the same: the system undergoes a smooth transition from the confining phase to the inverse confining phase as $\theta$ goes from $0$ to $\pi$, and this transition becomes weakened as temperature increases.

\section{Summary and future study} \label{sec:summary}
In this paper, we have revisited the lattice formulation of the bosonized Schwinger model of Bender et al. in 1985~\cite{Bender:1984qg} and shed new light on it as a method for circumventing the sign problem.
After conducting a comprehensive review of their accomplishment, we have verified the formulation by reproducing the analytical chiral condensate at $m = 0$ and also
by comparing our numerical results at $m \neq 0$ with previous numerical studies.
As an application, we have studied confining properties in the Schwinger model at finite temperature and $\theta$.
We have established the string tension in the $(T, \theta)$ plane for the first time and revealed the confining properties in the Schwigner model quantitatively.
In particular, we found that the string tension is negative for noninteger probe charges around $\theta = \pi$ at low temperatures.

The present method has both advantages and drawbacks compared to spin Hamiltonian approaches.
Because of the fast convergence to the continuum limit and low numerical cost,
no continuum nor infinite spatial length extrapolation is needed in practice, as was demonstrated in this paper.
Also, thermal expectation values can be very easily calculated in the method.
On the other hand, expectation values at exactly zero temperature cannot be obtained.
Moreover, it seems very hard to achieve the precision of the tensor network method at zero temperature. 
Thus, these two approaches complement each other and would promote further understanding of the Schwinger model.
In particular, investigating the fate of CP symmetry at finite temperature would be an attractive future study, which can be pursued using the present method.

As another future study, the application of the lattice formulation to other models would be intriguing.
The work of Bender et al. has revealed that it is possible to formulate a bosonized fermionic model on a lattice even with a nontrivial interaction, at least in the case of the Schwinger model.
An important question that arises here is the feasibility of applying the lattice formulation to other fermionic models in one spatial dimension, such as the multi-flavor Schwinger model, the gauged Thirring model, one-dimensional QCD, and so on.
Because the key formula~(\ref{eq:Coleman_latt}) is model-independent, and bosonization is a rather universal concept in one spatial dimension,
the lattice formulation is expected to be applicable to a wide variety of fermionic models.


\acknowledgments
The author was supported by a Grant-in-Aid 
for JSPS Fellows (Grant No.22KJ1662).
The numerical simulations have been carried out on Yukawa-21 at Yukawa Institute for Theoretical Physics (YITP), Kyoto University.
The author dedicates this paper to Akira Ohnishi, 
who was a professor at YITP and passed away on May 16, 2023.
This work unexpectedly emerged from collaborative research with him and Koichi Murase on a different subject.


\bibliographystyle{JHEP}
\bibliography{bosonization.bib}
\end{document}